\documentclass[prb,twocolumn,floatfix,amsmath,amssymb,
superscriptaddress]{revtex4}
\usepackage{graphicx}
\usepackage{dcolumn}
\usepackage{bm}
\begin{document}

\title{Many-body calculation of the spatial extent of the wave-function of a
    non-magnetic impurity in a d-wave high-temperature superconductor using the t--J model}
\author{Manuela Capello}
\affiliation{Laboratoire de Physique Th\'eorique, CNRS and Universit\'e 
de Toulouse, F-31062 Toulouse, France}

\author{Didier Poilblanc}
\affiliation{Institute of Theoretical Physics, Ecole Polytechnique F\'ed\'erale de Lausanne, CH-1015 Lausanne, Switzerland}
\affiliation{Laboratoire de Physique Th\'eorique, CNRS and Universit\'e 
de Toulouse, F-31062 Toulouse, France}

\date{\today}
\begin{abstract}

Scanning tunneling microscopy (STM) by providing images of the effects of individual zinc-impurities in cuprate superconductors
with unprecedented atomic-resolution offers a stringent test to models
of correlated fermions for high-temperature superconductors.
Using a t-J model supplemented by Variational Monte Carlo many-body techniques, the spatial dependence of the hole density 
and of the valence bond and superconducting pairing amplitudes 
around the impurity are computed.  A cross-shaped four-fold symmetric structure very similar to the observed STM observation is found, giving strong credit to the model.

\end{abstract}


\maketitle

{\it Introduction --}
Cuprates superconductors can be considered as doped two-dimensional (2D) Mott insulators where electronic correlations play a 
dominant role~\cite{Zha88,Sac03}. A number of exotic properties such as the pseudo-gap behavior reflect the complexity of the system. 
In a pioneering work,
Anderson proposed the Resonating Valence Bond (RVB) Mott insulator as the relevant underlying {\it parent} state~\cite{And87} from which 
gapless d-wave superconductivity naturally emerges under doping.
Within this scenario, the pseudo-gap naturally emerges as the energy scale associated to the formation of singlet electron pairs via the nearest-neighbor antiferromagnetic (AF) exchange. A mean-field version of the RVB theory using a t--J model~\cite{Zha88}
could also explain a number of bulk experimental
observations~\cite{RVB}.

Local probes of correlated materials with atomic resolution have recently become possible thanks to 
Scanning Tunneling Microscopy (STM)~\cite{Fis2007} which provided unprecedented high-resolution maps of the surface of 
some under-doped cuprate superconductors~\cite{Valla}. The measured space-resolved doped-hole charge density in the SC regime of Na-CCOC and Dy-Bi2212 cuprates revealed stripe patterns~\cite{Koh07}.
This discovery naturally raises the question
whether such inhomogeneities are induced by impurities
or whether they are intrinsic as the bulk
static charge and spin stripe orders detected 
in Nd-LSCO~\cite{ndlsco} and LBCO~\cite{lbco} cuprates at doping $\delta\sim 1/8$.

Substituting a single impurity atom for a copper atom indeed strongly affects its surrounding region.  Therefore, it can serve as a local fine probe, providing important insights about the properties of the correlated medium itself~\cite{Sca96}. Imaging the effects of individual zinc impurity atoms on superconducting
Bi2212 performed by STM~\cite{Pan00} showed clear real-space modulations which can be confronted to theoretical modeling. 
In other words, such observations offer a new stringent test to models
of correlated fermions for high-temperature superconductors.
It has been argued that a number of bulk properties of these
materials can be
explained within the correlated t--J model~\cite{dagotto}.
However, local real-space responses have not yet been calculated
reliably since, due to the short
superconducting coherence length, a fully many-body approach 
is needed. 
In this Letter, we have carried out such a program
of (i) computing within a many-body numerical technique
the response induced by the  introduction of a spinless impurity site within
a bulk two-dimensional t--J model and (ii) confronting the
theoretical results to the experimental observations
to validate or invalidate the model.

\begin{figure}
\begin{center}
\includegraphics[width=0.25\textwidth]{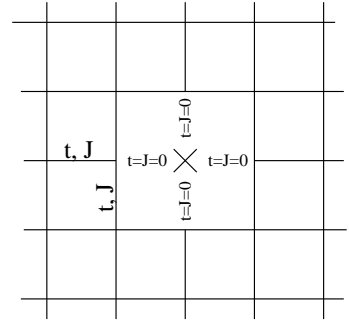}
\caption{\label{fig:model}
Schematic representation of the impurity model. The cross corresponds to the impurity site $i_0$. The parameters $t$ and $J$
are set to zero on the four bonds connected to the impurity site.}
\end{center}
\end{figure}

The $t-J$ Hamiltonian on a square lattice reads, 
\begin{equation}\label{Htj}
H_{t-J}=-\sum_{\langle i,j\rangle \sigma} 
t_{ij} (c^\dagger_{i\sigma} c_{j\sigma}+ h.c.)+  
\sum_{\langle i,j\rangle} J_{ij} S_i\cdot S_j \, .
\end{equation}
A zinc impurity is in the same $2+$ oxidation state 
as the copper ion it is substituted for, so that it does not introduce  extra charge. 
However, in contract to the copper S=1/2 ion, the Zn$^{2+}$ ion is in a spin-singlet state, inert magnetically. Hence, one can use a simple description~: on the four bonds connected to the impurity site we set $t_{ij}=0$ and $J_{ij}=0$ as shown on Fig.~\ref{fig:model}. On all the other bonds,
we set $t_{ij}$ and $J_{ij}$ to the same values $t$ and $J$, respectively. Although completely local such a "boundary" is expected to lead to a {\it spatially-extended}  perturbation strongly affecting  GS properties.

{\it Description of method --}
Variational fully-projected fermionic wave-functions~\cite{VMC}  (WF)
are known to incorporate very satisfactorily correlation effects
of the t-J model.  
We have extended them to finite periodic $L\times L$ clusters containing a single impurity (in practice, $L=8$ and $L=16$).
A Variational Monte Carlo (VMC) scheme~\cite{VMC} is used to
(i) realize a thorough optimization over the variational parameters~\cite{VMC2} (see below)
and (ii) to calculate the ground-state (GS) physical observables. The presence of the impurity on site $i_0$
modifies the Hilbert space, i.e.
$c^\dagger_{i_0\sigma}|\Psi\rangle=0$ and $c_{i_0\sigma}|\Psi\rangle=0$, 
where $|\Psi\rangle$ is the GS of the system.
In our Monte Carlo variational scheme an ``impurity projector'' 
${\cal{P}}_{i_0}=(1-n_{i_0\uparrow})(1-n_{i_0\downarrow})$ is inserted, and the impurity variational wavefunction is defined as
$|\Psi_{\rm VMC}\rangle={\cal{P}}_g {\cal{P}}_{i_0}|D\rangle$,
where ${\cal{P}}_g$ is the usual Gutzwiller projector enforcing the constraint of no-double occupancy on the remaining \hbox{$L^2 -1$} sites.

Motivated by the success of the RVB theory to explain bulk properties~\cite{RVB} the mean-field determinant $|D\rangle$
is chosen to be the ground-state of a mean-field Hamiltonian of standard BCS-type,
\begin{eqnarray}\label{eqHmf}
H_{\rm MF}=\sum_{\langle i,j\rangle \sigma} 
(\chi_{ij} c^\dagger_{i\sigma} c_{j\sigma}+ h.c.)+ \nonumber \\ 
+\sum_{\langle i,j\rangle} 
(\Delta_{ij} c^\dagger_{i\uparrow} c^\dagger_{j\downarrow}+ h.c.)
+ \mu\sum_{i\sigma} n_{i\sigma}  \, ,
\end{eqnarray}
defined on {\em all} of the $L\times L$ sites, 
including the $i_0$ site (always occupied by a hole).
We optimize {\it all} different 
non-equivalent bonds around the impurity, starting from an initial guess
respecting or not the square lattice symmetry around $i_0$. 
Since in principle the Hamiltonian $C_{4v}$ symmetry around $i_0$ 
(see Figure~\ref{fig:model}) could be spontaneously broken, we have performed a number of preliminary tests on small $8\times 8$ lattices,
choosing the initial RVB bonds pattern with lower symmetries like e.g. 
$C_{4}$ or $C_{2v}$ symmetries (the later allowing the formation of a domain wall).
We have found that the full $C_{4v}$ symmetry is systematically restored at the variational minimum.
Therefore, to reduce the number of variational parameters and 
gain accuracy, the $C_{4v}$ symmetry has been enforced on our largest $16\times 16$ cluster. 
As expected, all optimized WFs are found to show 
opposite signs of $\Delta_{ij}$ on any site-sharing vertical and horizontal bonds, hence reflecting the expected
{\it orbital d-wave character} of the superconducting order.
Lastly, we note that allowing {\it finite} values
of the parameters $\Delta_{ij}$ and $\chi_{ij}$ on the four bonds connected to the impurity is also important to gain energy 
as shown on Table I.
The lowest-energy state (II) is obtained for a full optimization of the $\Delta_{ij}$ and $\chi_{ij}$ parameters over {\it all bonds}.
Typically, $\chi_{ij}$ has a significant magnitude on the four bonds connected to the impurity site.
Moreover, for decreasing doping,
a sizable amplitude of $\Delta_{ij}$ also appears on the later bonds.

\begin{table}[htbp]
\caption{\label{tabenergy} Variational energy per site (in units of $t$)
for different projected WFs for the
$t-J$ model at doping 1/8 ($N_h=8$), for $t/J=3$ and a $8\times8$ cluster with an impurity. (I) $|D(8\times8)\rangle$ is optimized
        fixing $\Delta_{ij}=0$, $\chi_{ij}=0$  around the impurity and
         $\mu_{i_0}=0$. Everywhere else $\Delta_{ij}$ optimized 
         and $\chi_{ij}=1$; (II) $|D(8\times8)\rangle$ is fully optimized with all possible $\Delta_{ij}$ and $\chi_{ij}$ (including the impurity bonds).
         The (total) energy difference between (I) and (II) is $\sim 0.3 t$
         per impurity.}
\begin{tabular}{ccc}
\hline \hline
$|D\rangle$: & (I)  & (II)  \\
 $E_{VMC} [t]$:      & -0.41493(5) & -0.41968(5)\\
\hline 
\end{tabular}
\end{table}

{\it Results on $16\times 16$ clusters --}
We now turn to the VMC calculations on the 16$\times$16 
cluster with periodic-boundary conditions, assuming a physical
value of $t/J=3$. Here, we consider a
physical "core"  8$\times$8 region centered around the impurity (i.e. of the same size as our previous small cluster), where we impose a $C_{4v}$ symmetry around the impurity site. Outside, we assume a 
uniform d-wave superconducting background (bg) whose parameters $\chi_{i,i+{\hat x}}=\chi_{i,i+{\hat y}}=\chi_{\rm bg}$ and
$\Delta_{i,i+{\hat x}}=-\Delta_{i,i+{\hat y}}=\Delta_{\rm bg}$ are optimized simultaneously. This enables to reduce significantly the boundary effects and is justified since the spatial extension of the effect of the impurity rarely exceeds the assumed size of the core.

\begin{figure}[htbp]
\begin{center}
\unitlength=0.01\textwidth
\begin{picture}(50,65)
\put(12,35){\includegraphics*[width=0.32\textwidth]{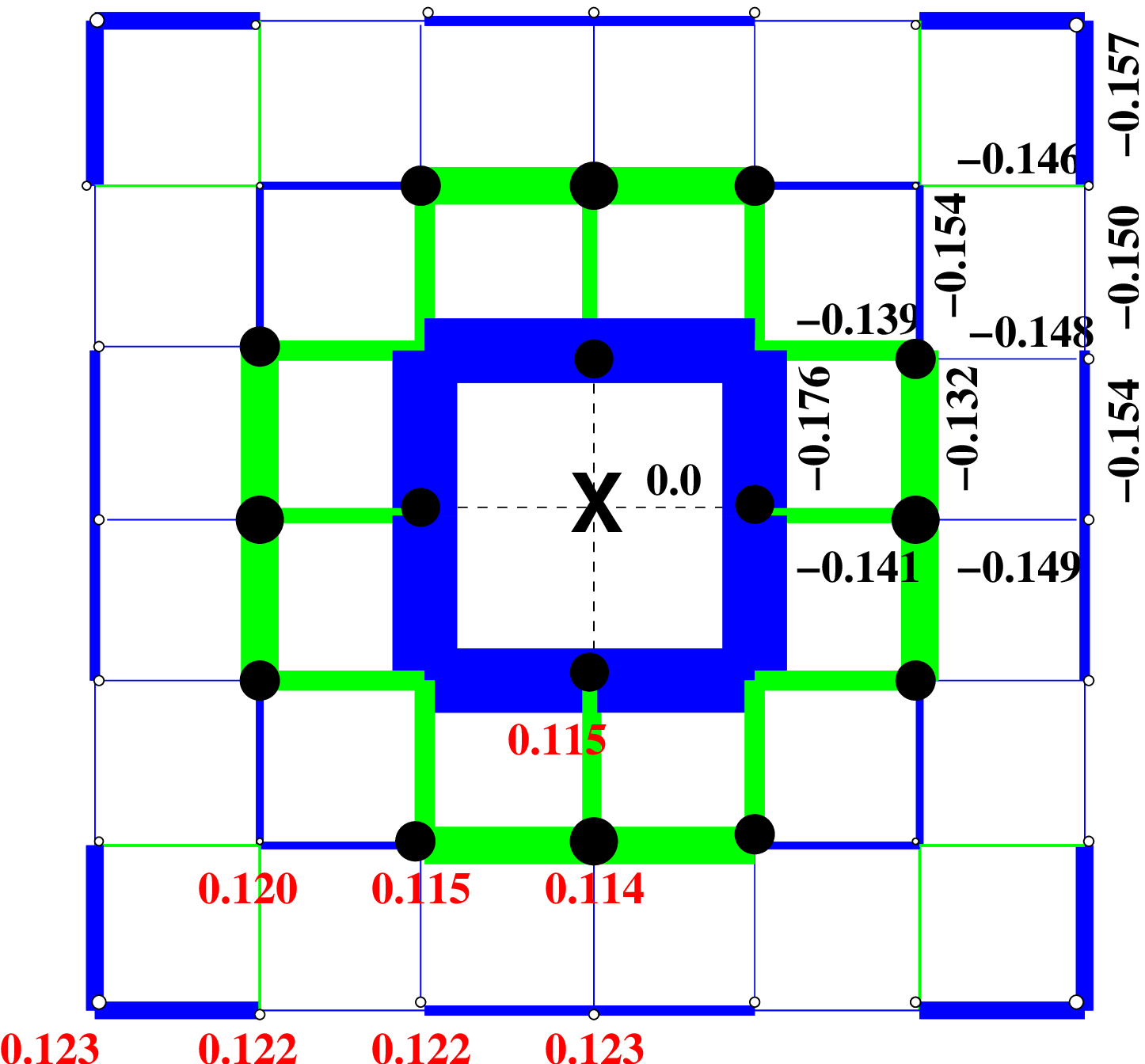}}
\put(12,0){\includegraphics*[width=0.32\textwidth]{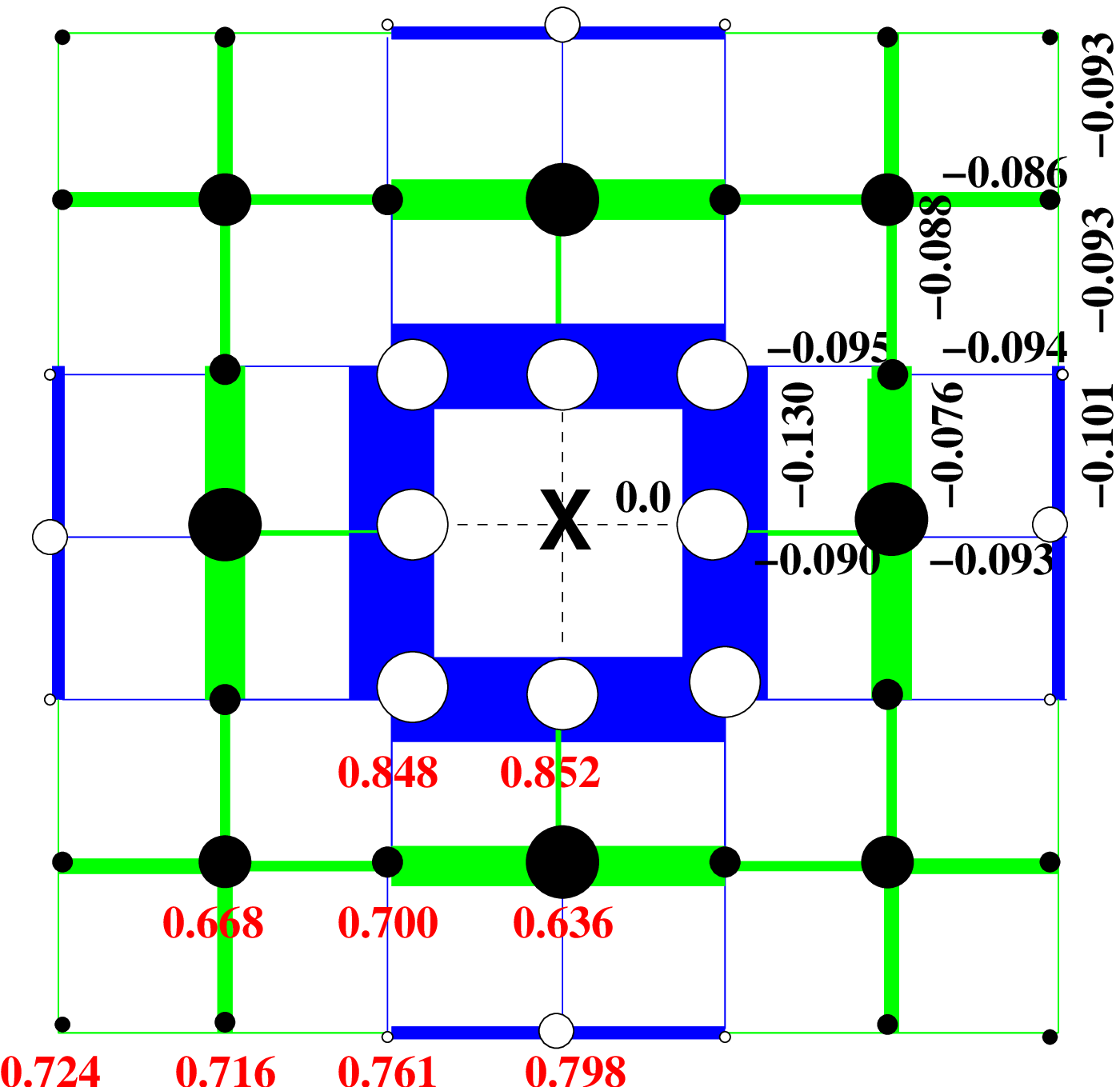}}
\put(2,10){ {\large (b)} }
\put(2,45){ {\large (a)} }
\end{picture}
\caption{\label{fig:hole_densities}
(Color online.) VMC results for the GS 
on-site hole densities (circles) and kinetic bond amplitudes (colored segments)
obtained on a $16\times16$ cluster.
Only the central region around the impurity is shown. 
Diameters of circles and widths of segments scale 
with the {\it absolute value} of the {\it relative} differences w.r.t. the impurity-free homogenous state at the same
$\delta_{\rm ave}$ hole density
(whose reference values are 
estimated by interpolating pure clusters with available flanking hole densities).
Higher (lower) hole densities and bond magnitudes
w.r.t. the homogeneous case  are shown by open (filled) circles and blue (green) bonds respectively. 
For completeness, we also show on the plot the (bare) numerical values of the non-equivalent sites/bonds. 
(a) and (b) corresponds to 32 ($\delta_{\rm ave}\simeq 0.1255$) and 20 
($\delta_{\rm ave}\simeq 0.0784$) doped holes giving rise, for an homogeneous background, to $E_{\rm kin}^{\rm homog}/t=-0.1487$ and $E_{\rm kin}^{\rm homog}/t= -0.0941$ per bond, respectively.}
\end{center}
\end{figure}

\begin{figure}[htbp]
\begin{center}
\unitlength=0.01\textwidth
\begin{picture}(50,60)
\put(6,24){\includegraphics*[width=0.38\textwidth]{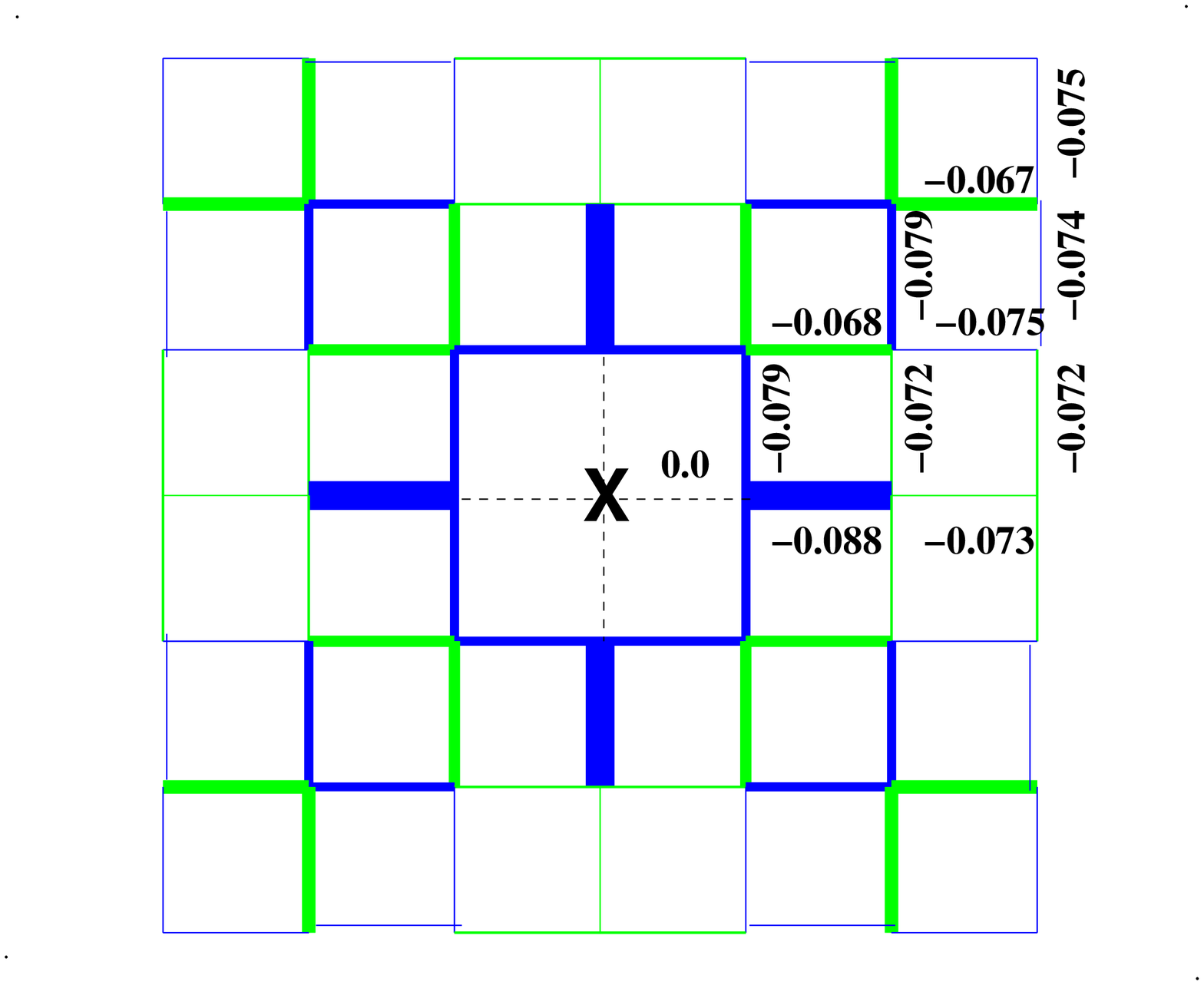}}
\put(10,-4){\includegraphics*[width=0.35\textwidth]{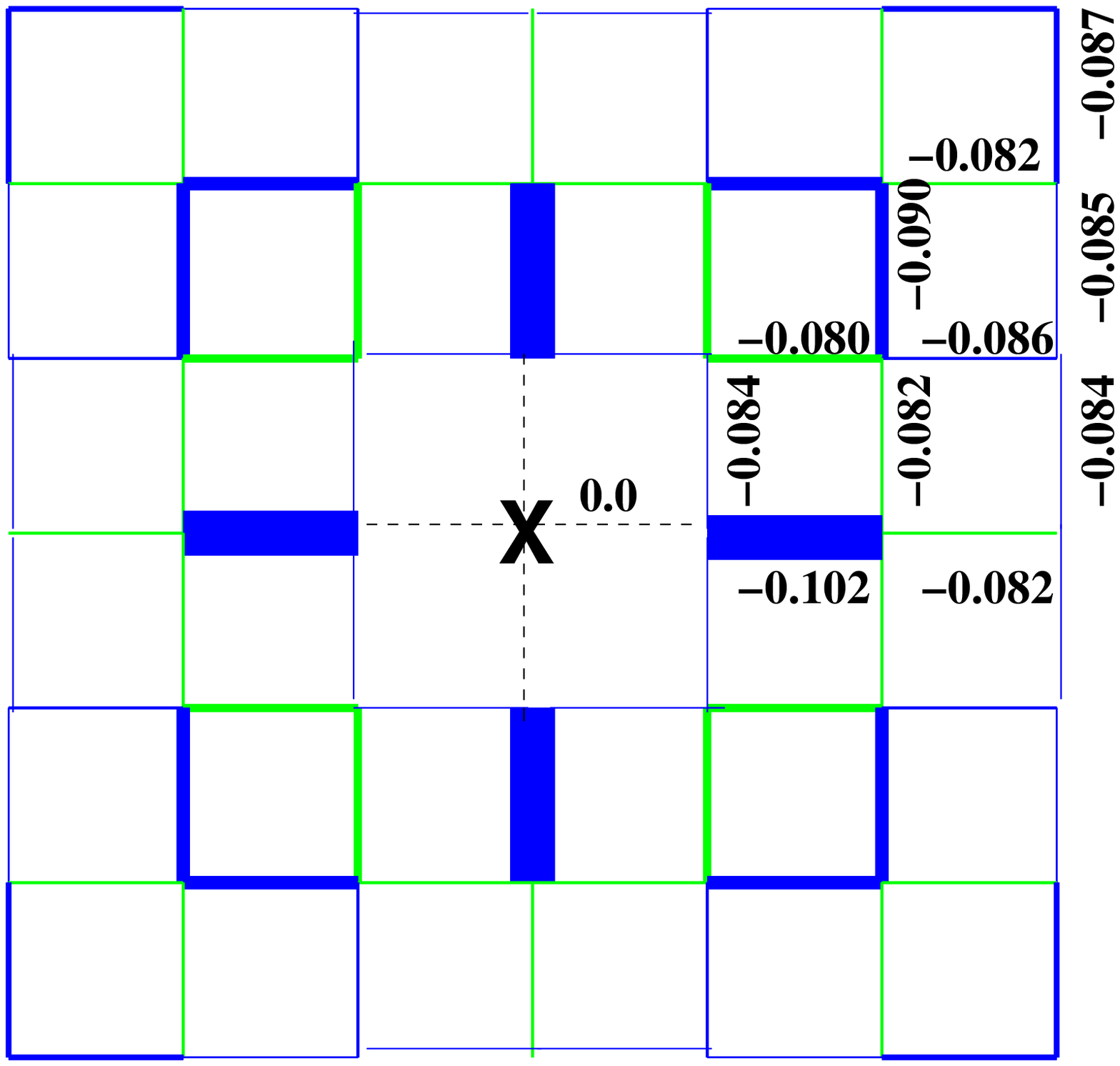}}
\put(2,10){ {\large (b)} }
\put(2,45){ {\large (a)} }
\end{picture}
\caption{\label{fig:magnetic}
(Color online.) VMC results for the GS magnetic bond amplitudes obtained on a $16\times16$ cluster.
Same conventions and parameters as in Fig.\protect\ref{fig:hole_densities}.
(a) and (b) corresponds to 32 ($\delta_{\rm ave}\simeq 0.1255$) and 20 
($\delta_{\rm ave}\simeq 0.0784$) doped holes giving rise for an homogeneous background to $E_{\rm mag}^{\rm homog}/J=-0.074$ and $E_{\rm mag}^{\rm homog}/J= -0.084$ per bond, respectively.}
\end{center}
\end{figure}

The spacial distribution 
of the local hole density $\big< c_{i\sigma}c_{i\sigma}^\dagger\big>$, the bond hole kinetic amplitudes  $K_{ij}=\big< (c_{j\sigma}c_{i\sigma}^\dagger+{\rm h.c.})\big>$ and the magnetic VB amplitudes $S_{ij}=\big< {\bf S}_{i}\cdot
{\bf S}_{j}\big>$ are shown in Fig.~\ref{fig:hole_densities}(a) and in Fig.~\ref{fig:magnetic}(a), respectively, for doping 1/8. 
Here, and throughout the paper, we only show the $6\times 6$ central region 
exhibiting the largest modulations.
It turns out that the variational parameters $\Delta_{ij}$ are suppressed on the four bonds connected to 
the impurity. This is compensated by an increase in $\Delta_{ij}$ and
hence of $S_{ij}$ on the
neighboring bonds, forming a cross-like structure (see thick
blue bonds of Fig.~\ref{fig:magnetic}(a)). Due to similarities
with work done in a
somewhat different context~\cite{Kaul2008}, we shall refer to these bonds as the four  ``dimer bonds''.
These bonds are characterized by a simultaneous hole deficiency
and a large gain in the magnetic energy (which can reach more than $40\%$), hence signaling a tendency towards singlet crystallization around the impurity. The distribution of $K_{ij}$ in Fig.~\ref{fig:hole_densities} shows also a remarkably strong modulation 
around the impurity.

\begin{figure}
\begin{center}
\unitlength=0.01\textwidth
\begin{picture}(50,60)
\put(13,30){\includegraphics*[width=0.29\textwidth]{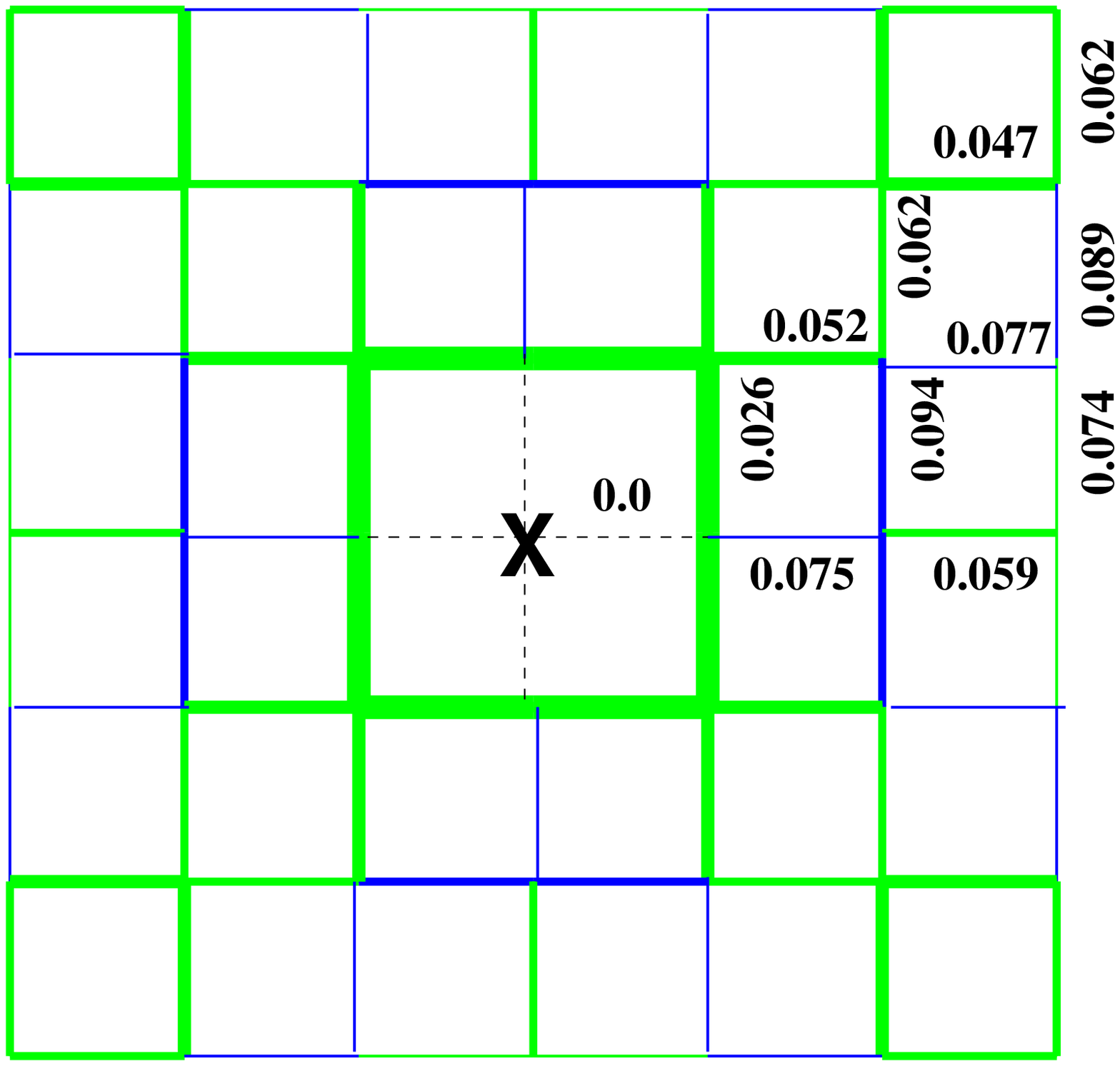}}
\put(9,-4){\includegraphics*[width=0.36\textwidth]{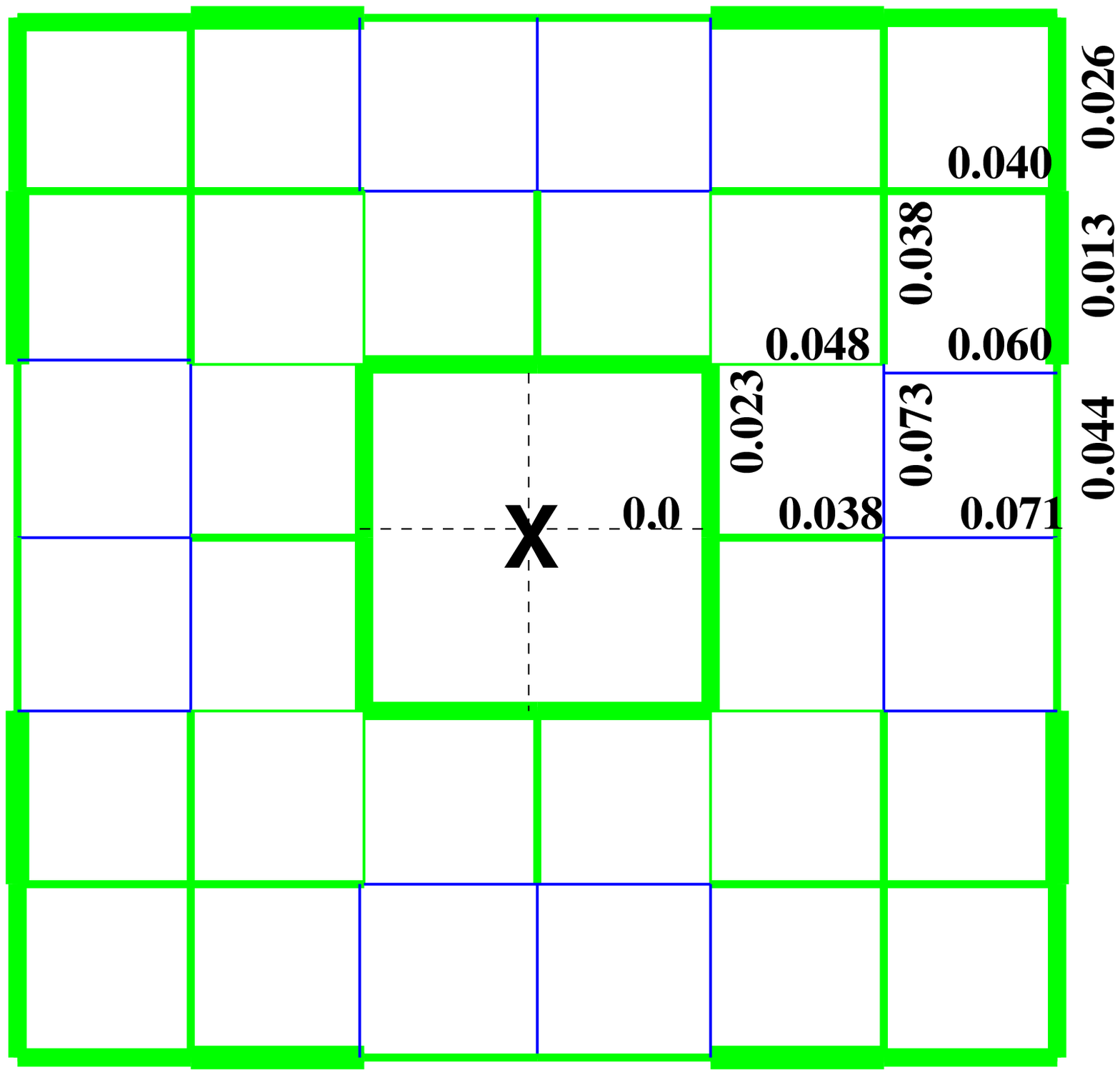}}
\put(2,10){ {\large (b)} }
\put(2,45){ {\large (a)} }
\end{picture}
\caption{\label{fig:pairing}
(Color online.) VMC results for the pairing bond amplitudes $\bar{\Delta}_{ij}$ obtained on a $16\times16$ cluster.
Same conventions and parameters as in Fig.\protect\ref{fig:hole_densities}.  
(a) and (b) corresponds to 32 
($\delta_{\rm ave}\simeq 0.1255$) and 20 ($\delta_{\rm ave}\simeq 0.0784$) doped holes giving  rise, for an homogeneous background, to $\Delta_{\rm bg}\simeq 0.0787$ and $\Delta_{\rm bg}\simeq 0.0626$, respectively.}
\end{center}
\end{figure}

Superconducting properties of RVB states
are characterized by 
the singlet-pair correlations at distance ${\bf r}$,
$\langle \Psi_{\rm VMC}| \tilde\Delta^\dag_{\bf s+r} \tilde\Delta_{\bf s} |\Psi_{\rm VMC} \rangle /
\langle \Psi_{\rm VMC}|\Psi_{\rm VMC} \rangle$, 
where the operator 
$\tilde\Delta^\dag_{\bf s} = c^\dag_{i({\bf s}),\uparrow} c^\dag_{j({\bf s+{\hat a}}),\downarrow} -
c^\dag_{i({\bf s}),\downarrow} c^\dag_{j({\bf s+{\hat a}}),\uparrow}$
creates a singlet pair of electrons on the bond between 
locations $\bf s$ and $\bf s+{\hat a}$ on the lattice,
$\bf {\hat a}$
being the unit vector that specifies the bond direction (along $x$ or $y$).
On the 8$\times 8$ cluster, we have computed pairing correlations between separate bonds for increasing bond separation.
However, at the largest distance available on this cluster, the correlations have not completely reached saturation.
To get a better estimation 
of the superconducting order parameter we have considered
the $16\times 16$ cluster and computed the pairing amplitudes 
$\bar{\Delta}_{ij}$ for all bonds $(i,j)$ within the "core" region,
\begin{equation}
\bar{\Delta}_{i({\bf s}),j({\bf s+{\hat a}})}=\frac{\langle {\tilde \Delta}_{\bf s} {\tilde \Delta}_{\rm bg}\rangle}
{\sqrt{\langle {\tilde \Delta}_{\rm bg} {\tilde \Delta}_{\rm bg}\rangle}} \, ,
\end{equation}
where ${\tilde \Delta}_{\rm bg}$ is a pair operator on the most remote bond in the homogeneous background. 
As shown in Fig.~\ref{fig:pairing}(a), pairing is enhanced on the dimer bonds, and is depleted
around the impurity, where holes are less present.

The 16$\times$16 cluster also allows to reduce the doping content, e.g. to 20 holes, going further into the under-doped region.
Interestingly, the hole distribution around the defect is very sensitive to
the doping ratio. Indeed, for doping around $12.5\%$  (32 holes) we found that holes
are slightly repelled from the bonds around the impurity.
In contrast, for $7.8\%$ doping, holes tend to concentrate more around the impurity site as shown in Fig.~\ref{fig:hole_densities}(b).
The variational pairing $\Delta_{ij}$ becomes stronger on the impurity bonds suggesting the formation of a "hole pair" with the impurity {\it empty} site. The corresponding real space modulations of $S_{ij}$
and $\bar{\Delta}_{ij}$ are shown in Figs.~\ref{fig:magnetic}(b) and \ref{fig:pairing}(b).  

{\it Discussions --}
Let us now compare our findings to prior theoretical approaches.
The first investigation of a single impurity immersed in
a correlated host 
has been performed using Lanczos exact diagonalization of small clusters. A calculation of the local density of states~\cite{Lanc94} revealed bound-states (of different orbital symmetries) in which a
mobile hole is trapped by the induced impurity potential.
Here, a unique mobile hole was assumed in the cluster,
hence preventing real bulk pairing and giving rise to a very small doping $\sim 5\%$ in the surrounding region.
Although our VMC calculations are done in a different physical range (and on much larger clusters), we find, for decreasing doping, the emergence of excess hole density around the impurity, which possibly could be consistent with  a bound-state formation when $\delta\rightarrow 0$.

Metlitski and Sachdev~\cite{Met2008} have introduced a theory of valence bond solid (VBS) correlations near a single impurity in a square lattice antiferromagnet. When the system is close to a quantum transition from a magnetically ordered N\'eel state to a spin-gap state with long-range VBS order, a missing spin
gives rise to a VBS pinwheel (or "vortex") around the impurity~\cite{Kaul2008}. To compare with these predictions we have 
computed the VBS 
order parameter of Eq.(5) in Ref.~\protect\onlinecite{Kaul2008}.
However, despite many similarities (e.g. crystallization of dimer bonds in the vicinity of the impurity), the vortex structure is not recognizable in our simulation. We hypothesize that (i)
our system is probably not close enough to the critical point 
assumed in Ref.~\protect\onlinecite{Met2008,Kaul2008} and/or (ii) the VBS region develops differently in a d-wave 
RVB than in an AF background.

Lastly, our results are compared to the experimental STM observations
around a Zn impurity in a Bi2212 cuprate superconductor shown in Ref.~\onlinecite{Pan00}. First, we point out that GS properties have been calculated here while Ref.~\onlinecite{Pan00} reports spectral properties. However, since equal-time and frequency-dependent 
quantities are related via a simple frequency integration up to a physical cutoff (see e.g. Ref.~\onlinecite{Koh07} for derivation of the local hole-charge distribution from space-resolved tunneling spectra) both sets of data should reveal similarities. Indeed, like in the experiments, the patterns we found show clearly a cross-shaped symmetric structure 
providing evidence that (i) the theoretical modeling of the impurity as an empty site and (ii) the use of the strongly correlated t--J model to describe the bulk high-Tc superconductor are realistic. 

\begin{acknowledgments}
We acknowledge support from the French Research Council
(ANR). D.P. also thanks S. Sachdev for insightful discussions.

\end{acknowledgments}


\begin{thebibliography}{00}
\bibitem{note1} Strictly speaking, for $N_h$ mobile holes in the cluster, the average doping of the surrounding medium is $\delta_{\rm ave}=N_h/(L^2-1)$, higher than the nominal doping 
$\delta=N_h/L^2$.
\end{thebibliography}
\end{document}